\documentclass[aps,prl,amsmath,nofootinbib,twocolumn]{revtex4}
\usepackage[]{graphicx}
\usepackage{amsmath}
\usepackage{amsfonts}
\usepackage{amssymb}
\usepackage{bm}

\usepackage{amsmath,amssymb,amsfonts,dcolumn,color,graphicx,graphics,latexsym,placeins,epsfig}

\def\R{{\mathcal R}}

\newcommand{\be}{\begin{equation}}
\newcommand{\ee}{\end{equation}}
\newcommand{\bea}{\begin{eqnarray}}
\newcommand{\eea}{\end{eqnarray}}

\newcommand{\reff}{{\mbox{\tiny ref}}}
\newcommand{\eqn}[1]{(\ref{#1})}
\usepackage{braket}
\begin{document}

\title{One spectrum to rule them all?}

\author{Alexander Gallego Cadavid$^{1,3}$, Antonio Enea Romano$^{1,2,4}$ 
}
\affiliation
{
${}^{1}$Instituto de F\'isica, Universidad de Antioquia, A.A.1226, Medell\'in, Colombia\\
${}^{2}${Theoretical Physics Department, CERN, CH-1211 Geneva 23, Switzerland}\\
${}^{3}$Escuela de F\'isica, Universidad Industrial de Santander, Bucaramanga, Colombia\\
${}^{4}$ICRANet, Piazza della Repubblica 10, I--65122 Pescara, Italy \\
}
\begin{abstract}
We show that in absence of entropy or effective anisotropic stress the freedom in the choice of the initial energy scale of inflation implies the existence of an infinite family of dual slow-roll parameters histories which can produce the same spectrum of comoving curvature perturbations.  
This  implies that in general there is no one-to-one correspondence between the spectrum and higher order correlation functions.
We give some  numerical examples of expansion histories corresponding to different initial energy scales, with the same spectrum of curvature perturbations, the same squeezed limit bispectrum, in agreement with the squeezed limit consistency condition, but with different bispectra in other configurations and different spectra of primordial gravitational waves. The combined analysis of data from future CMB and gravitational wave experiments could allow to distinguish between dual models.
\end{abstract}

\maketitle

\section{Introduction}
Cosmic microwave background (CMB) observations are compatible with an approximately scale invariant primordial curvature perturbations power spectrum \cite{wmapfmr,Adam:2015rua,Ade:2015xua,Ade:2015lrj, Chung:2005hn}. Higher order correlation functions allow to distinguish between inflationary models producing the same spectrum, and it is important to investigate the relation between them. In this paper we consider what are the general conditions for different single field inflationary models to produce the same spectrum of comoving curvature perturbations, showing that the freedom in choosing the initial energy scale of inflation gives an infinite class of dual models. Our results are model independent  and can be applied  to any model for which the anisotropy and  entropy effects are negligible, including multifields models in the regime in which isocurvature perturbations are not important. We give some examples of dual models with the same spectrum and different bispectra of curvature perturbations, and also different spectra of primordial gravitational waves.
In particular we show that there can exist dual models producing a scale invariant spectrum of curvature perturbation but completely different bispectra, and different primordial gravitational waves spectrum.

We also consider the case of models with local features in the expansion history, which are related to other types of features such as features of the inflaton potential ~\cite{Starobinsky:1992ts, Adams2, Adams:2001vc,  Gariazzo:2014dla, Gariazzo:2015qea, Hazra:2014jwa,  Hazra:2016fkm, Cadavid:2015iya, GallegoCadavid:2016wcz, GallegoCadavid:2015bsn, GallegoCadavid:2017pol, Palma:2014hra, pp, Arroja:2011yu, Chen:2011zf, Hazra:2014goa, Martin:2014kja, Romano:2014kla, Starobinsky:1998mj}, and  can provide a better fit to  observational data at the scales  where the spectrum shows some deviations from  power law~\cite{constraints2,Adams2, Adams:2001vc,Gariazzo:2015qea, Gariazzo:2014dla,Hazra:2014jwa, Hunt:2013bha, Joy:2007na, Joy:2008qd, Hazra:2016fkm}.
Finally we consider the implications of the existence of dual models for consistency relations between the spectrum and bispectrum of primordial curvature perturbations~\cite{Palma:2014hra, Romano:2016gop, Chung:2005hn, Mooij:2015cxa, Creminelli:2012ed}.

\section{Primordial curvature perturbations}\label{pctp}
The study of primordial scalar perturbations is attained by expanding perturbatively the action with respect to the background Friedmann-Lemaitre-Robertson-Walker (FLRW) solution~\cite{m,Cheung:2007st}. In absence of entropy perturbations or effective anisotropic stress, the second-order action for scalar perturbations in the comoving gauge is
\bea\label{eq:s2}
 S_2^{\mathcal{R}_{c}} = M_{Pl}^2 \int dt d^3x\left[\frac{a^3 \epsilon}{c_s^2} \dot{\mathcal{R}_{c}^2}-a\epsilon(\partial \mathcal{R}_{c})^2 \right] \,,
\eea
where $ M_{Pl}$ is the reduced Planck mass, $a$ is the scale factor, $c_s$ is the scalar sound speed, $\mathcal{R}_c$ is the curvature perturbation on comoving slices, and we denote the derivatives with respect to time with dots. Throughout this paper we use units of $\hbar=c=1$. For the slow-roll parameters we use the following definitions
\be \label{eq:slowroll}
  \epsilon \equiv -\frac{\dot H}{H^2} \, , \quad \eta \equiv \frac{\dot \epsilon}{\epsilon H},
\ee
where $H \equiv \dot a/a$ is the Hubble parameter. The Euler-Lagrange equations for this action give
\begin{equation}\label{pert}
 \frac{\partial}{\partial t}\left(\frac{a^3 \epsilon}{c_s^2} \frac{\partial \mathcal{R}_{c}}{\partial t}\right)- 
a\epsilon\delta^{ij} \frac{\partial^2 \mathcal{R}_{c}}{\partial x^i\partial x^j}=0.
\end{equation}
This equation is quite general and can also be derived from the perturbed Einstein's equations. It can in fact be shown that Eq. \eqn{pert} is satisfied by the curvature perturbations produced by an arbitrary physical system described by an effective energy momentum tensor with no entropy perturbations or effective anisotropic stress. This is for example the case of a quintessence. Given the wide class of physical systems to which this equation can be applied, also our results are quite general.

Taking the Fourier transform of the previous equation we obtain the equation of motion for the primordial curvature perturbation 
\begin{equation}\label{eq:cpe}
  \mathcal{R}_{c}''(k) + 2 \frac{z'}{z} \mathcal{R}_{c}'(k) + c^2_s k^2 \mathcal{R}_{c}(k) = 0,
\end{equation}
where $z\equiv a\sqrt{2 \epsilon}/c_s$~\cite{Langlois:2010xc}, $k$ is the comoving wave number, and primes denote derivatives with respect to conformal time $d\tau \equiv dt/a$. 
As initial conditions for Eq.~\eqn{eq:cpe} we take the Bunch-Davies vacuum \cite{Bunch:1978yq, Noumi:2014zqa, Langlois:2010xc} 
\bea 
\label{eq:ic1}
\mathcal{R}_c(\tau_i,k) &=& \frac{v_i(k, c_s)}{z_i  M_{Pl}} \\ 
\label{eq:ic2}
\mathcal{R}'_c(\tau_i,k) &=& \frac{1}{ M_{Pl}} \Bigl( \frac{v'_i(k, c_s)}{z_i}-\frac{v_i(k, c_s)}{z_i} \frac{z'_i}{z_i} \Bigr)\, ,
\eea
where from now on we denote quantities evaluated at initial time $t_i$ or $\tau_i$ by the subscript $i$ and
\begin{equation}
  v(k,c_s)=\frac{e^{-\mathrm{i} c_s k  \tau}}{\sqrt{2c_s k }}\left(1-\frac{\mathrm{i}}{c_s k \tau}\right)\,.
\end{equation}

\subsection{Primordial spectrum of curvature perturbations}\label{pscp}
The two-point correlation function of primordial curvature perturbations is given 
by~\cite{Ade:2015lrj,Planck:2013jfk}
\begin{equation}
 \Braket{ \hat{\mathcal{R}}_c(\vec{k}_1, \tau_e) \hat{\mathcal{R}}_c(\vec{k}_2, \tau_e) } \equiv (2\pi)^3 
\frac{2\pi^2}{k^3} P_{\mathcal{R}_c}(k) \delta^{(3)}(\vec{k}_1+\vec{k}_2) \, ,
\end{equation}
where $\tau_e$ is the exit horizon time and the power spectrum of primordial curvature perturbations is defined as 
\bea
P_{\mathcal{R}_{c}}(k) \equiv \frac{2k^3}{(2\pi)^2}|\mathcal{R}_{c}(k, \tau_e)|^2.
\eea

\subsection{Primordial bispectrum of curvature perturbations}\label{pbcp}
The three-point correlation function is given by~\citep{m,xc}
\bea
\Braket{ \hat {\mathcal{R}_c}(\tau_e,\vec{k}_1) \hat {\mathcal{R}_c}(\tau_e,\vec{k}_2)  \hat {\mathcal{R}_c}(\tau_e,\vec{k}_3) } \\ \notag
\equiv (2\pi)^3 B_{\mathcal{R}_c}(k_1,k_2,k_3) \delta^{(3)}(\vec{k}_1+\vec{k}_2+\vec{k}_3) ,
\eea
where
\bea \label{b} \nonumber
  && B_{\mathcal{R}_c}(k_1,k_2,k_3)= 2 \Im\Bigl[ \mathcal{R}_c(\tau_e,k_1) \mathcal{R}_c(\tau_e,k_2)\mathcal{R}_c(\tau_e,k_3)\\ \nonumber
  && \int^{\tau_e}_{-\infty} d\tau \eta(\tau) \epsilon(\tau) a^2(\tau) \biggl( 2\mathcal{R}_c^*(\tau,k_1) \mathcal{R}_c'{}^*(\tau,k_2)\mathcal{R}_c'{}^*(\tau,k_3) \\ 
 &&  - k^2_1 \mathcal{R}_c^*(\tau,k_1)\mathcal{R}_c^*(\tau,k_2)\mathcal{R}_c^*(\tau,k_3) + \mbox{ perms.} \biggr)  \Bigr]\, ,
\eea
is the bispectrum and ``perms." means the other two permutations of $k_1,k_2$ and $k_3$.

In this work we will study the degeneracy in the bispectrum using the usual $f_{NL}$~\citep{hann} quantity which in the case of our definition of the spectrum takes the form
\bea\label{FNL}
  f_{NL}(k_1,k_2,k_3)=\frac{10}{3}\frac{(k_1 k_2 k_3)^3}{(2\pi)^4} \\
\frac{B_{\mathcal{R}_c}(k_1,k_2,k_3)}{P_{\mathcal{R}_c}^2(k_1)P_{\mathcal{R}_c}^2(k_2)k_3^3 +  \mbox{perms.} } \, \notag .
\eea

\subsection{Primordial gravitational waves}
The primordial tensor  perturbations satisfy the equation
~\cite{Planck:2013jfk,Ade:2015lrj}
\bea \label{eq:tpe}
h''_k+2\frac{z'_{\gamma}}{z_\gamma} h'_k+ c^2_\gamma k^2 h_k=0 \, ,
\eea 
where $c_{\gamma}$ is the tensor sound speed~\cite{Noumi:2014zqa, Cai:2016ldn},
$z_\gamma \equiv a /c_\gamma$, and now the initial conditions for the tensor modes are
\bea 
\label{eq:ic1h}
h(\tau_i,k) &=& \frac{\sqrt{2} v_i(k, c_{\gamma})}{z_{i,\gamma}  M_{Pl}} \\ 
\label{eq:ic2h}
h'(\tau_i,k) &=& \frac{\sqrt{2}}{ M_{Pl}} \Bigl( \frac{v'_i(k, c_{\gamma})}{z_{i,\gamma}}-\frac{v_i(k, c_{\gamma})}{z_{i,\gamma}} \frac{z'_{i,\gamma}}{z_{i,\gamma}} \Bigr)\, .
\eea

\subsection{Primordial spectrum of gravitational waves}\label{psgw}
The two-point correlation function of tensor perturbations is 
\begin{equation}
 \Braket{ \hat{h}^{s}(\vec{k}_1, \tau_e) \hat{h}^{s'} (\vec{k}_2, \tau_e) } \equiv (2\pi)^3 
\frac{\pi^2}{2k^3} P_{h}(k) \delta^{(3)}(\vec{k}_1+\vec{k}_2) \delta_{s s'} \, ,
\end{equation}
where $s=\pm$ is the helicity  index and the power spectrum of primordial tensor perturbations is defined as
\bea
P_{h}(k) \equiv \frac{2k^3}{\pi^2}|h_k(\tau_e)|^2\,.
\eea

\section{How many expansion histories  can give the same spectrum?}\label{dcp}
As can be seen from Eq.~\eqn{eq:cpe} the evolution of the primordial curvature perturbation $\mathcal{R}_c$ is entirely determined by $c_s$, the 
background quantity $z'/z$, and the initial conditions in Eqs.~\eqn{eq:ic1} and~\eqn{eq:ic2}, which are in terms of $z_i$ and $z_i'$. 

For a given spectrum how much freedom is left in specifying the slow-roll parameters history or equivalently the expansion history?
Mathematically this question corresponds to ask under what conditions we can obtain the same solution for the curvature perturbation equation, and the answer is that this is possible as long as $c_s$, the coefficient $z'/z$ of the equation, and the initial conditions are the same. 

After choosing  a reference function $z_\reff$  we can find the family of expansion histories such that the corresponding equation of curvature perturbation has the same coefficient, i.e.
\be
\label{eq:zpz}
\frac{z'}{z} = \frac{z'{}_\reff }{z_\reff}\, . 
\ee
A general solution for $z$ is
\be
z(\tau) = C z_\reff \,,
\ee
where $C$ is an arbitrary integration constant. From now on we will assume that $c_s$ is the same for the dual and the reference models.
In order for the spectrum to be the same also the initial conditions have to be the same, which implies that 
\be
z (\tau) = z_\reff (\tau) \,.
\ee
This means that different dual models can have the same spectrum of primordial curvature perturbations as long as their $z(\tau)$ is the same at all times.
Note that this condition is not enough to ensure also $z_{\gamma}=z_{\gamma ~ref}$, implying that the spectrum of tensor perturbations of dual models will be different, as we will show later in different cases.

The interesting fact is that models with the same $z$ can have different slow-roll parameters, or equivalently expansion histories. 
Note that the coefficients of the curvature perturbations equation depend only on the sound speed and the scale factor, and for this reason, in the spirit of the effective field theory of inflation \cite{Cheung:2007st}, we can make a completely model independent analysis focusing on them without having to specify the Lagrangian of the model. This model independent analysis can be carried out by using the relation between $z$ and the scale factor in terms of the cosmic time $t$
\begin{equation}\label{eq:z}
z=\frac{a\sqrt{2\epsilon}}{c_s}= \frac{1}{c_s} \sqrt{2\Biggl(a^2 - \frac{ a^3 \ddot{a}}{\dot{a}^2}\Biggl)}.
\end{equation}
For a given functio $z_\reff$ the scale factor evolution is not uniquely determined.
From the above relation we get in fact a  second-order differential equation for the scale factor
\begin{equation}\label{eq:diffa}
a^2 - \frac{ a^3 \ddot{a}}{\dot{a}^2} = \frac{1}{2} z_\reff^2 c_s^2\,.
\end{equation}

 The initial value of the scale factor has no physical importance since it can always be arbitrarily fixed, but the initial condition for the first time derivative is physically important since it corresponds to consider background histories with different \textit{initial Hubble} parameters $H_i$, and consequently different  \textit{initial energy scales}. We will parametrize this difference in the initial energy scale with the dimensionless quantity $\delta H=H_i/H_{\reff, i}$.
 
This freedom in choosing $H_i$ while keeping the same evolution of the function $z(\tau)$ is the origin of the existence of an infinite set of expansion histories producing the same spectrum of curvature perturbations, which was found in some specific class of models in \cite{GallegoCadavid:2017bzb}.



Note that we have derived the conditions to obtain dual models with exactly the same spectrum for $\R_c$, while in the past \cite{Wands:1998yp,Boyle:2004gv} the conditions to have an exactly scale invariant spectrum were investigated. 
For scale invariance the condition is weaker and it is enough to study the Sasaki-Mukhanov variable~\cite{Arroja:2011yu} $u=a \delta \phi=-\mbox{sign}(\dot{\phi}) \R_c z M_{Pl}$ which satisfy the equation
\bea\label{pertu}
u_k''+ \left(c_s^2 k^2-\frac{z''}{z}\right) u_k&=&0 \,.
\eea

In this case the condition is that $z''/z$ has to be the same, giving $z=z_{ref}\int d\tau z_{ref}^{-2}$, while to get the same spectrum the condition is $z=z_{ref}$. Our results are consistent with those obtained in \cite{Wands:1998yp}, since the sign of the initial condition $\dot{a}_i$  corresponds to expanding and contracting background solutions. We have shown that in general there is an infinite class of expansion histories (contracting or inflationary), which in the context of the study of spectra with features has some important phenomenological implications in regard to the relation with the bispectra or other higher-order correlation functions. In general in fact models with the same spectrum could have different bispectra.

\begin{figure}
  \includegraphics[scale=0.6]{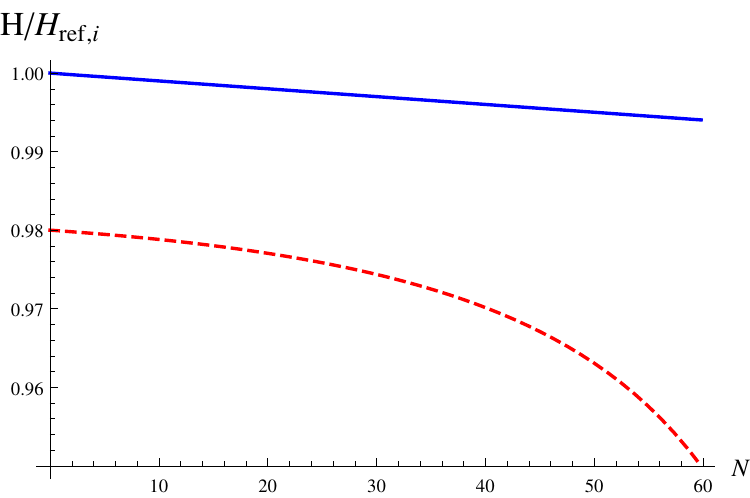}
\caption{The numerically computed Hubble parameter $H$ of the dual model obtained from Eq.~\eqn{eq:diffa} (dashed red) for $\delta H= 0.98$ and $H_\reff$ (blue) corresponding to the model in Eq.~\eqn{eq:aref1} are plotted as functions of the number of $e$-folds $N$. }
\label{fig:H}
\end{figure}

\begin{figure}
 \begin{minipage}{.45\textwidth}
  \includegraphics[scale=0.6]{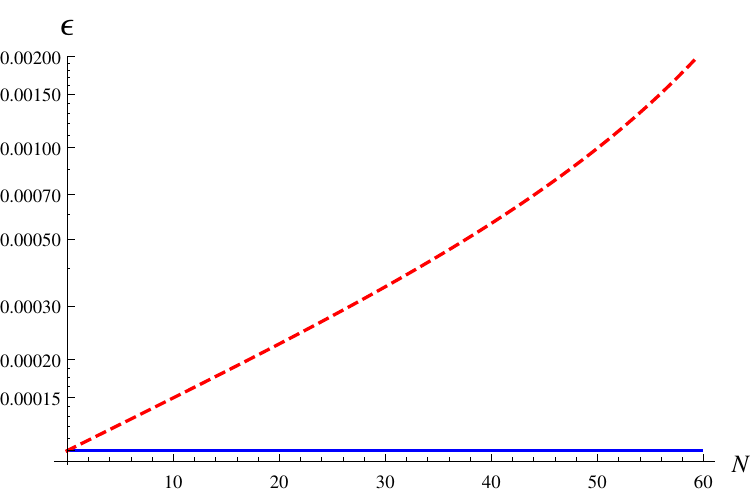}
  \end{minipage}
 \begin{minipage}{.45\textwidth}
  \includegraphics[scale=0.6]{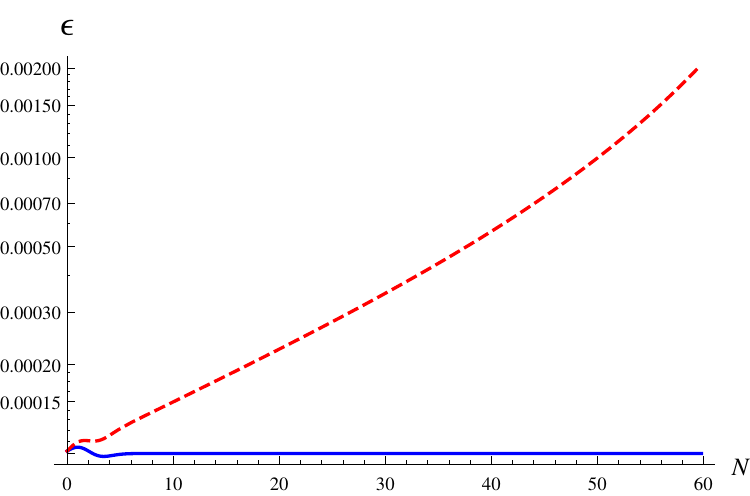}
 \end{minipage}
  \caption{The numerically computed slow-roll parameter $\epsilon$ for the dual model (dashed red) for $\delta H=0.98$ and $\epsilon_\reff$ (blue) are plotted as functions of the number of $e$-folds $N$. The top plot shows the results obtained for the featureless model defined in Eq.~\eqn{eq:aref1}, and the bottom plot for the model with features given in Eq.~\eqn{eq:aref2}.
}
\label{fig:sreps}
\end{figure}

\begin{figure}
 \begin{minipage}{.45\textwidth}
  \includegraphics[scale=0.6]{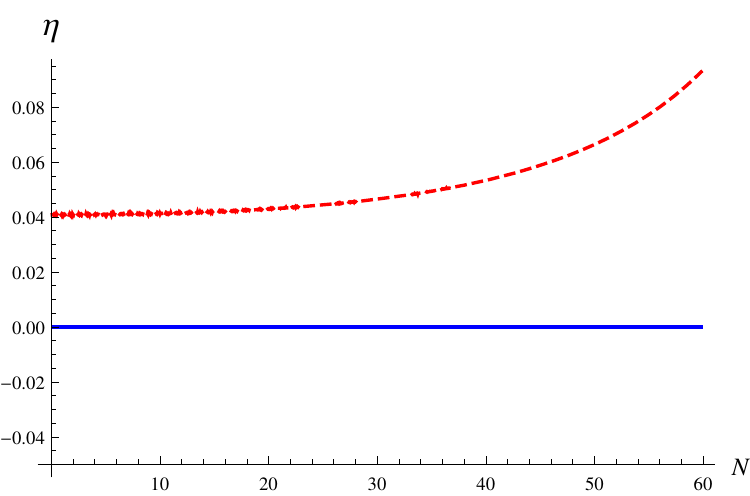}
  \end{minipage}
 \begin{minipage}{.45\textwidth}
  \includegraphics[scale=0.6]{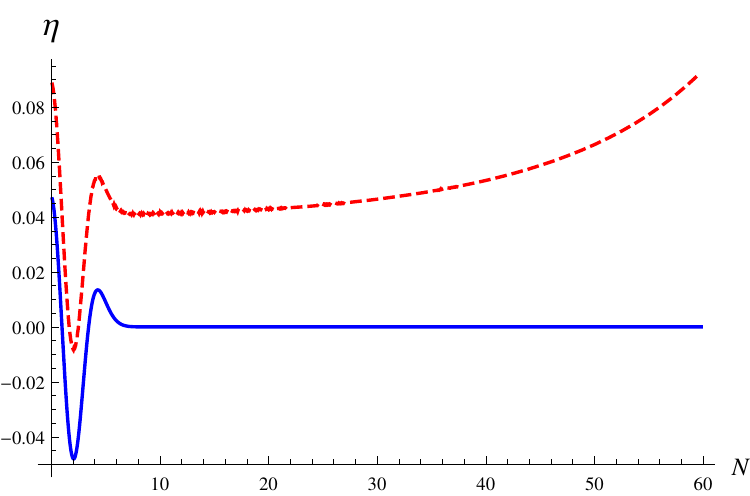}
 \end{minipage}
  \caption{The numerically computed slow-roll parameter $\eta$ for the dual model (dashed red) for $\delta H=0.98$ and $\eta_\reff$ (blue) are plotted as functions of the number of $e$-folds $N$. The top plot shows the results obtained for the featureless model defined in Eq.~\eqn{eq:aref1}, and the bottom plot for the model with features given in Eq.~\eqn{eq:aref2}.
}
\label{fig:sreta}
\end{figure}

\section{An example of a dual model}\label{edm} 
Let us consider a model with constant slow-roll parameter $\epsilon$, corresponding to the scale factor
\be \label{eq:aref1}
a_{\reff}(t)= \Bigr(1+ \epsilon_{\mbox{\tiny c}} H_{\reff, i} t \Bigl)^{1/\epsilon_{\mbox{\tiny c}}},
\ee
where $H_{\reff, i}=\dot{a}_{\reff, i}/a_{\reff, i}$, $\epsilon_c$ is a constant, and we have chosen the initial condition $a(t_i=0)=1$. For this model analytic expressions can be derived  for the background quantities and the spectrum of primordial curvature and tensor perturbations, facilitating the comparison with the dual models obtained numerically, following these steps
\begin{itemize}
    \item solve equation Eq.~\eqn{eq:diffa} for the dual $a$, where $z_\reff$ is calculated using $a_\reff$ from Eq.~\eqn{eq:aref1}
    \item different dual expansion histories giving the same $z_\reff$ correspond to different initial conditions for the dual scale factor $\dot{a}_{i}$ or equivalently different initial value of the dual Hubble parameter $H_{i}$. 
    \item from the dual $a$ we compute the corresponding slow-roll parameters histories, and from them the spectra of curvature and tensor perturbations and the bispectrum of curvature perturbations.
\end{itemize}

As shown in the next section the spectra of curvature perturbations of the dual models  are exactly the same but the spectra of tensor perturbations and the bispectra of curvature perturbations are different.

In order to obtain an analytic formula for the scalar spectrum we first write Eq.~\eqn{eq:aref1} in terms of conformal time
\be
a_\reff (\tau) = - H_{\reff,i} (1-\epsilon_c  \tau )^{1/(\epsilon_c-1)} ,
\ee
such that
\be \label{azppz}
\frac{z_\reff''(\tau)}{z_\reff(\tau)} = \frac{2-\epsilon_c}{ (1-\epsilon_c)^2  \tau ^2} .
\ee
From this expression, and using the Sasaki-Mukhanov variable in Eq.~\eqn{pertu}, we can obtain an exact solution when $c_s=1$ for the curvature perturbation~\cite{ Wang:2013eqj, 2009arXiv0902.1529K}
\be\label{aR}
\R_c(\tau,k)=\frac{\sqrt{-k \tau }}{z M_{Pl}} \Bigl[C_1 J_\nu{(-k \tau )} + C_2 Y_\nu{(-k \tau )} \Bigr],
\ee
where $J_\nu(z)$ and $Y_\nu(z)$ are the Bessel functions of the first and second kind, respectively, $\nu= (\epsilon_c-3)/(2 (\epsilon_c-1))$, and $C_1$ and $C_2$ are given by the initial conditions in  Eqs.~\eqn{eq:ic1} and~\eqn{eq:ic2}. A similar expression is found for the tensor modes $h_k$~\cite{Wang:2013eqj, 2009arXiv0902.1529K}.

\begin{figure}
 \begin{minipage}{.45\textwidth}
  \includegraphics[scale=0.6]{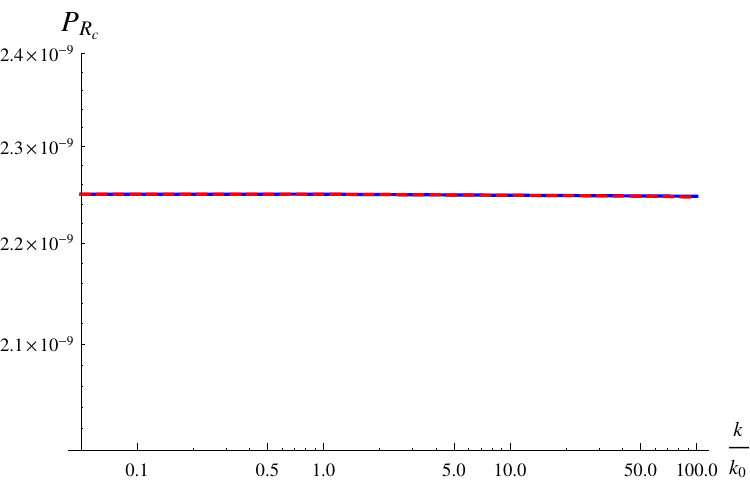}
  \end{minipage}
 \begin{minipage}{.45\textwidth}
  \includegraphics[scale=0.6]{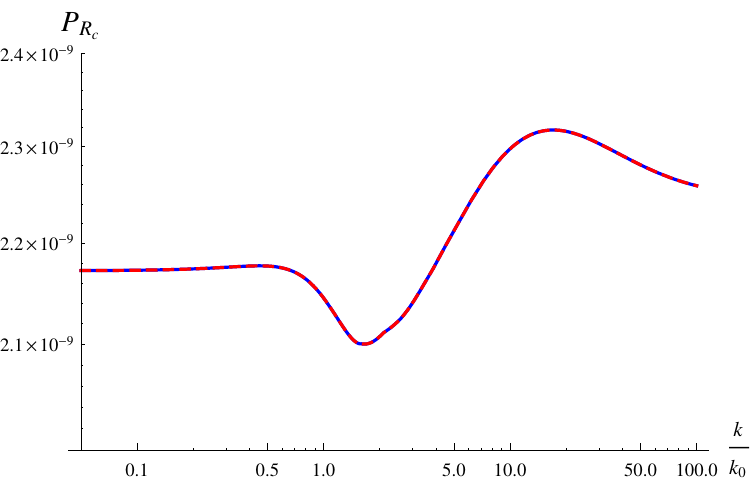}
 \end{minipage}
  \caption{The numerically computed spectrum of primordial curvature perturbations $P_{\mathcal{R}_c}$ for the dual model (dashed red) for $\delta H=0.98$ and for the reference model (blue) are plotted as functions of the scale $k$. The top plot shows the results obtained for the reference models corresponding to the featureless case defined in Eq.~\eqn{eq:aref1}, and the bottom plot for the case with features given in Eq.~\eqn{eq:aref2}. For the featureless model in the top the spectrum is computed analytically  Eq.~\eqn{aR}.
  }
\label{fig:Pplot}
\end{figure}

\begin{figure}
 \includegraphics[scale=0.6]{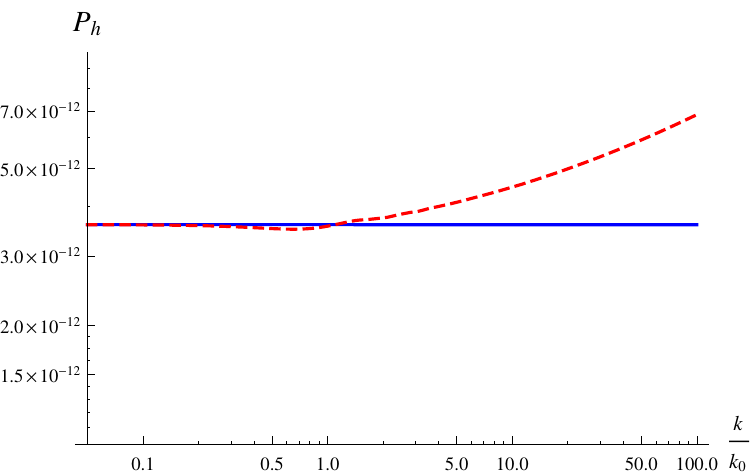}
  \caption{The numerically computed spectrum of primordial tensor perturbations $P_{h}$ for the dual model (dashed red) for $\delta H=0.98$ and for the reference model (blue) are plotted as functions of the scale $k$. The top plot shows the results obtained for the featureless model defined in Eq.~\eqn{eq:aref1}, and the bottom plot for the model with features given in Eq.~\eqn{eq:aref2}.
  }
\label{fig:Phplot}
\end{figure}

\begin{figure}
 \begin{minipage}{.45\textwidth}
  \includegraphics[scale=0.5]{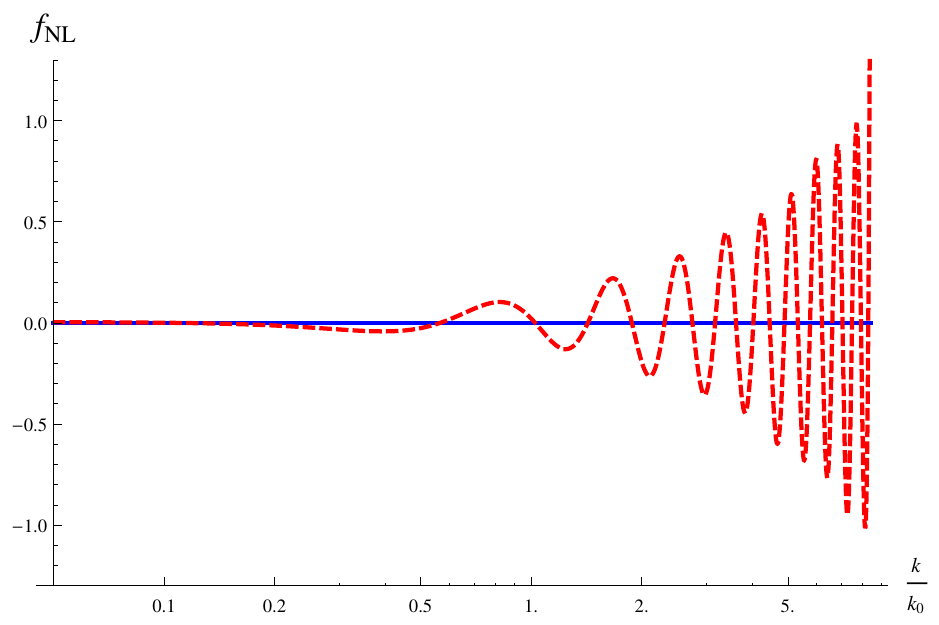}
  \end{minipage}
 \begin{minipage}{.45\textwidth}
  \includegraphics[scale=0.6]{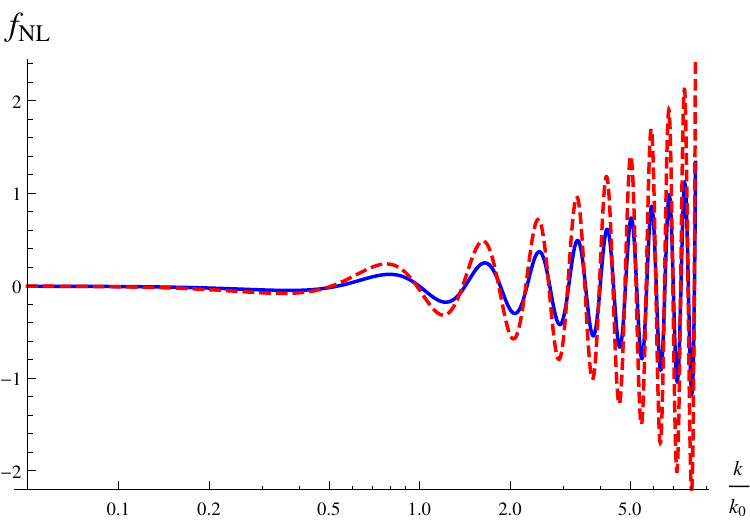}
 \end{minipage}
  \caption{The numerically computed $f_{NL}$ functions in the squeezed configuration ($k_1=k \ll k_2=k_3=1000 k_0$) for the dual model (dashed red) for $\delta H=0.98$ and for the reference model (blue) are plotted as functions of the scale $k$. The top plot shows the results obtained for the featureless model defined in Eq.~\eqn{eq:aref1}, and the bottom plot for the model with features given in Eq.~\eqn{eq:aref2}. In the squeezed limit ($k \rightarrow 0$) the bispectra are approximately the same, in agreement with the SCC.}
\label{fig:FNLsl}
\end{figure}

\begin{figure}
 \begin{minipage}{.45\textwidth}
  \includegraphics[scale=0.6]{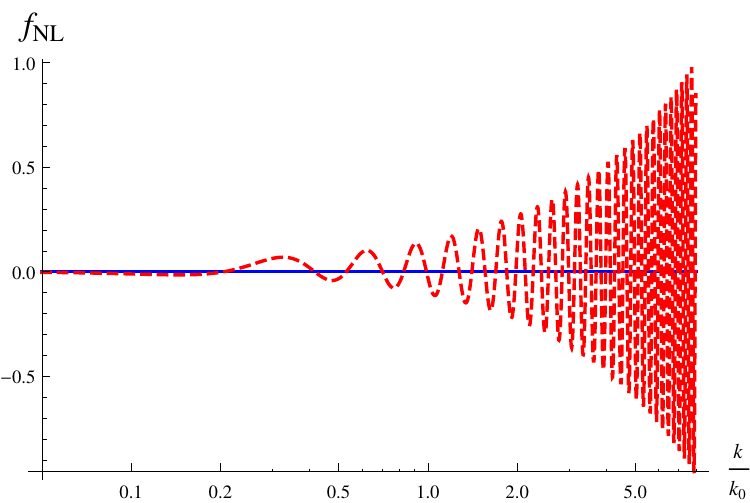}
  \end{minipage}
 \begin{minipage}{.45\textwidth}
  \includegraphics[scale=0.6]{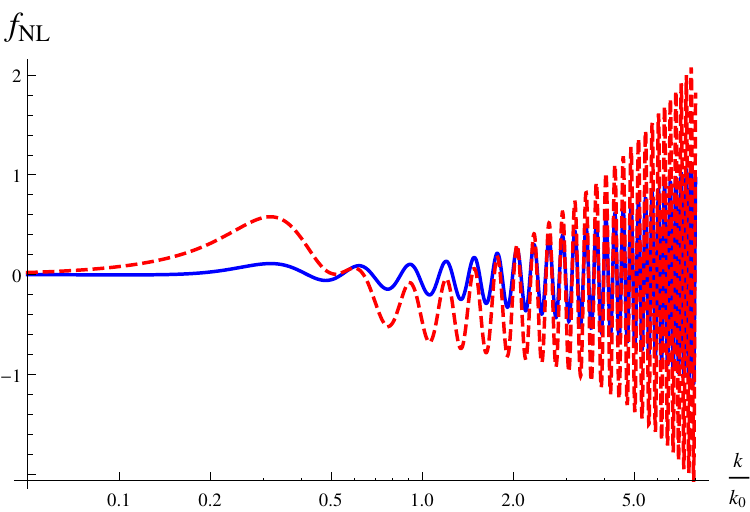}
 \end{minipage}
  \caption{The numerically computed $f_{NL}$ functions in the equilateral configuration for the dual model (dashed red) for $\delta H=0.98$ and for the reference model (blue) are plotted as functions of the scale $k$. The top plot shows the results obtained for the featureless model defined in Eq.~\eqn{eq:aref1}, and the bottom plot for the model with features given in Eq.~\eqn{eq:aref2}.
  }
\label{fig:FNLel}
\end{figure}

We show our results in Figs.~\ref{fig:H} -~\ref{fig:FNLel}, where $N \equiv \ln[a/a_i]$ is the number of $e$-folds  after the beginning of inflation. The values of the parameters used in the figures are $H_{\reff,i}=4.22\times 10^{-6} M_{Pl}, \epsilon_c=10^{-4}$, $c_s=c_{\gamma}=1$, and $\delta H=0.98$.
In the plots we use the reference scale $k_0$, corresponding to modes exiting the horizon at time $t_0$.  
For the squeezed configuration $f_{NL}$ we use $k_1=k \ll k_2=k_3=1000~k_0$. It can be  seen that, even though the curvature spectrum is the same for dual models, they can be distinguished   at the bispectrum level. 

Since the coefficients of Eq.~\eqn{eq:tpe} depend on the scale factor in a form different than in the equation for comoving curvature perturbation, we expect that also the spectra of primordial gravitational waves of dual models will be different, as confirmed in Fig.~\ref{fig:Phplot}

\section{An example of dual models with features}\label{edmf} 
In order to understand the implications of the existence of dual models in presence of features ~\cite{Starobinsky:1992ts, Adams2, Adams:2001vc,  Gariazzo:2014dla, Gariazzo:2015qea, Mooij:2015cxa, Ashoorioon:2006wc, Hazra:2014jwa,  Hazra:2016fkm, Cadavid:2015iya, Romano:2014kla, GallegoCadavid:2016wcz, GallegoCadavid:2015bsn, GallegoCadavid:2017pol, Palma:2014hra, pp, Arroja:2011yu, Chen:2011zf, Hazra:2014goa, Martin:2014kja, Cai:2015xla, Starobinsky:1998mj} we consider a local ~\cite{Cadavid:2015iya, GallegoCadavid:2016wcz}  modification of the scale considered in the previous section, given by
\be \label{eq:aref2}
a_{\reff}(t)= \Bigr(1+ \epsilon_{\mbox{\tiny c}} H_{\reff, i} t \Bigl)^{1/\epsilon_{\mbox{\tiny c}}}  \Bigr[1+ \lambda e^{-(\frac{t-t_0}{\sigma})^2} \Bigl],
\ee
where $\lambda$ and $\sigma$ are parameters that control the magnitude and width of the feature, and $t_0$ is the feature time. The results are shown in the bottom panel of Figs.~\ref{fig:H} -~\ref{fig:FNLel}. As can be seen, even in the presence of features, the spectrum of curvature perturbations of dual models is exactly the same, while the tensor spectrum and the curvature bispectrum are different.

\begin{figure}
 \includegraphics[scale=0.65]{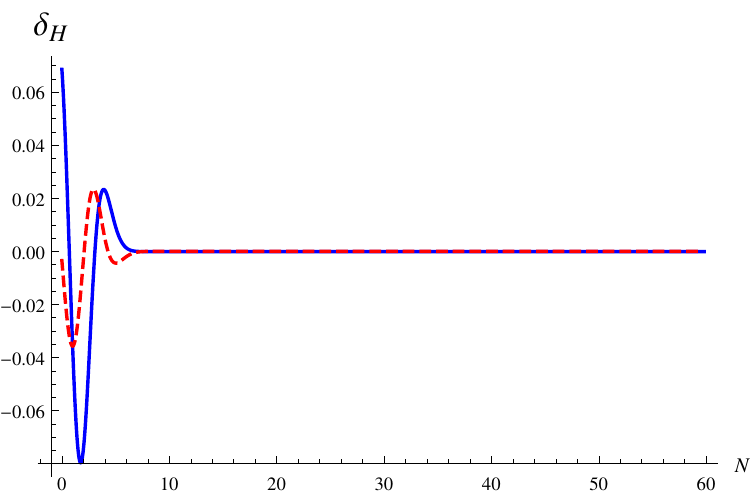}
 \caption{The numerically computed $\delta_H$  (blue) and its sudden change approximation given in  Eq.~\eqn{dhapp} (red)  are plotted for the reference model defined in Eq.~\eqn{eq:aref2} as a function of the number of $e$-folds. As can be seen the approximation is not accurate, implying a violation of the GCC derived under the assumption of its validity. This is confirmed by the difference among the bispectra of the dual  models as shown in Fig.(6-7). The SCC does not require the validity of the sudden change approximation, so the despite  the latter is not accurate,  the SCC is satisfied as shown by the fact that dual models  have the same squeezed limit ($k \rightarrow 0$) bispectra.}
\label{zppz}
\end{figure}

\section{Consistency conditions}

Attractor single field slow roll models satisfy the squeezed limit consistency condition (SCC)~\cite{m,Creminelli:2004yq,Romano:2016gop}
\bea
\label{mcc}
\lim_{k_1 \to 0} \Braket{ \hat{\mathcal{R}}_{\vec{k}_1} \hat{\mathcal{R}}_{\vec{k}_2} \hat{\mathcal{R}}_{\vec{k}_3} } =  
- (2 \pi)^3 \delta^3(\sum_i \vec k_i)(n_s-1) P_{k_1} P_{k_3},\nonumber
\eea
which holds also in presence of features~\cite{Adshead:2013zfa, Passaglia:2018afq}. 
Since  dual models have the same spectrum, it is expected that they should also have the same squeezed limit bispectrum, as shown in the low $k_1$ limit of the squeezed configuration in Fig. 6, confirming that they satisfy the SCC.

Under  the sudden change approximation~\cite{Palma:2014hra} 
\be
\frac{z''}{z} \equiv  2a^2H^2 \Bigl(1 +\frac{1}{2} \delta_H \Bigr) \approx 2a^2H^2 \Bigl(1-\frac{1}{2}\tau \eta' \Bigr)
\label{dhapp} 
\ee
another consistency condition for general bispectrum configurations (GCC)  has been derived  
\bea
f_{NL}  &\simeq&   \frac{5}{12} \frac{k_1 k_2 k_3}{k_1^3 + k_2^3 + k_3^3}  \left[ \frac{d^2}{d \ln k^2}  \frac{\Delta P_{\mathcal{R}_{c}}}{P_{\mathcal{R}_{c}}^0} (k)  \right]_{k = (k_1 + k_2 + k_3)/2}  \label{quasi-equi}  
\eea
where  $\Delta P_{\mathcal{R}_{c}} \equiv P_{\mathcal{R}_{c}}(k) - P_{\mathcal{R}_{c}}^0 $ is the difference between the featured power spectrum $P_{\mathcal{R}_{c}}(k)$ and its featureless counterpart $P_{\mathcal{R}_{c}}^0$. Note that the GCC is much stronger than the SCC because it relates the spectrum to the bispectrum in any configuration. The GCC only applies to models satisfying the sudden change approximation given in Eq.~(\ref{dhapp}), contrary to the SCC which holds for any attractor single field model.
In Fig.~\ref{zppz} we plot the numerically computed $\delta_H$, showing that Eq.~(\ref{dhapp}) is not  a good approximation for the models with features we consider. 

If always valid the GCC  would imply that models with the  same spectrum should also have the same bispectrum  \textit{in any configuration}, but as shown in Figs.~\ref{fig:FNLsl}-\ref{fig:FNLel}, this is not true for the dual models we constructed. This should not be surprising because for a general non-Gaussian field the bispectrum cannot be fully determined  by its spectrum.
In general there is in fact an infinite family  of dual slow-roll parameter histories which do not satisfy the sudden change approximation in Eq.~(\ref{dhapp}) and consequently violate  the GCC derived assuming it, while still satisfying the SCC, which does not rely on the validity of the sudden change approximation.

As a direct consequence for any given spectrum there is an infinite class of single field models  satisfying the SCC but violating the GCC.

\section{Conclusions}
We have shown that there is an infinite family of slow-roll parameters histories which can produce the same spectrum of comoving curvature perturbations. This degeneracy is related to the freedom in the choice of the initial conditions for the second order differential equation relating the coefficients of the curvature perturbation equations to the scale factor, and it corresponds to fixing the initial energy scale of inflation.
This freedom implies that in general there is no one-to-one correspondence between the spectrum and higher order correlation functions, unless some special conditions are satisfied by the slow-roll parameters. 
We have given  some examples of dual models  with the same spectrum, the same squeezed limit bispectrum, in agreement with the squeezed limit consistency condition, but different bispectra in other configurations, and different primordial gravitational wave spectra. A combined analysis of data from future CMB experiments such as the CMB-S4 \cite{Abazajian:2016yjj} and  space gravitational detectors such as the Laser Interferometer Space Antenna (LISA)~\cite{2017arXiv170200786A} and Evolved Laser Interferometer Space
Antenna (eLISA)~\cite{Seoane:2013qna}  could allow to distinguish  between different dual models.

\bibliography{Bibliography}

\begin{thebibliography}{10}

\bibitem{wmapfmr}
WMAP, C.~Bennett {\em et~al.},
\newblock Astrophys.J.Suppl. {\bf 208}, 20 (2013), arXiv:1212.5225.

\bibitem{Adam:2015rua}
Planck, R.~Adam {\em et~al.},
\newblock (2015), arXiv:1502.01582.

\bibitem{Ade:2015xua}
Planck, P.~A.~R. Ade {\em et~al.},
\newblock Astron. Astrophys. {\bf 594}, A13 (2016), arXiv:1502.01589.

\bibitem{Ade:2015lrj}
Planck, P.~A.~R. Ade {\em et~al.},
\newblock Astron. Astrophys. {\bf 594}, A20 (2016), arXiv:1502.02114.

\bibitem{Chung:2005hn}
D.~J. Chung and A.~Enea~Romano,
\newblock Phys.Rev. {\bf D73}, 103510 (2006), arXiv:astro-ph/0508411.

\bibitem{Starobinsky:1992ts}
A.~A. Starobinsky,
\newblock JETP Lett. {\bf 55}, 489 (1992).

\bibitem{Adams2}
J.~A. Adams, G.~G. Ross, and S.~Sarkar,
\newblock Nucl.Phys. {\bf B503}, 405 (1997), arXiv:hep-ph/9704286.

\bibitem{Adams:2001vc}
J.~A. Adams, B.~Cresswell, and R.~Easther,
\newblock Phys. Rev. {\bf D64}, 123514 (2001), arXiv:astro-ph/0102236.

\bibitem{Gariazzo:2014dla}
S.~Gariazzo, C.~Giunti, and M.~Laveder,
\newblock JCAP {\bf 1504}, 023 (2015), arXiv:1412.7405.

\bibitem{Gariazzo:2015qea}
S.~Gariazzo, L.~Lopez-Honorez, and O.~Mena,
\newblock Phys. Rev. {\bf D92}, 063510 (2015), arXiv:1506.05251.

\bibitem{Hazra:2014jwa}
D.~K. Hazra, A.~Shafieloo, and T.~Souradeep,
\newblock JCAP {\bf 1411}, 011 (2014), arXiv:1406.4827.

\bibitem{Hazra:2016fkm}
D.~K. Hazra, A.~Shafieloo, G.~F. Smoot, and A.~A. Starobinsky,
\newblock JCAP {\bf 1609}, 009 (2016), arXiv:1605.02106.

\bibitem{Cadavid:2015iya}
A.~G. Cadavid, A.~E. Romano, and S.~Gariazzo,
\newblock Eur. Phys. J. {\bf C76}, 385 (2016), arXiv:1508.05687.

\bibitem{GallegoCadavid:2016wcz}
A.~Gallego~Cadavid, A.~E. Romano, and S.~Gariazzo,
\newblock Eur. Phys. J. {\bf C77}, 242 (2017), arXiv:1612.03490.

\bibitem{GallegoCadavid:2015bsn}
A.~Gallego~Cadavid and A.~E. Romano,
\newblock Nucl. Part. Phys. Proc. {\bf 267-269}, 254 (2015).

\bibitem{GallegoCadavid:2017pol}
A.~Gallego~Cadavid,
\newblock J. Phys. Conf. Ser. {\bf 831}, 012003 (2017), arXiv:1703.04375.

\bibitem{Palma:2014hra}
G.~A. Palma,
\newblock JCAP {\bf 1504}, 035 (2015), arXiv:1412.5615.

\bibitem{pp}
A.~E. Romano and M.~Sasaki,
\newblock Phys.Rev. {\bf D78}, 103522 (2008), arXiv:0809.5142.

\bibitem{Arroja:2011yu}
F.~Arroja, A.~E. Romano, and M.~Sasaki,
\newblock Phys.Rev. {\bf D84}, 123503 (2011), arXiv:1106.5384.

\bibitem{Chen:2011zf}
X.~Chen,
\newblock JCAP {\bf 1201}, 038 (2012), arXiv:1104.1323.

\bibitem{Hazra:2014goa}
D.~K. Hazra, A.~Shafieloo, G.~F. Smoot, and A.~A. Starobinsky,
\newblock JCAP {\bf 1408}, 048 (2014), arXiv:1405.2012.

\bibitem{Martin:2014kja}
J.~Martin, L.~Sriramkumar, and D.~K. Hazra,
\newblock JCAP {\bf 1409}, 039 (2014), arXiv:1404.6093.

\bibitem{Romano:2014kla}
A.~G. Cadavid and A.~E. Romano,
\newblock Eur. Phys. J. {\bf C75}, 589 (2015), arXiv:1404.2985.

\bibitem{Starobinsky:1998mj}
A.~A. Starobinsky,
\newblock Grav.Cosmol. {\bf 4}, 88 (1998), arXiv:astro-ph/9811360.

\bibitem{constraints2}
D.~K. Hazra, M.~Aich, R.~K. Jain, L.~Sriramkumar, and T.~Souradeep,
\newblock JCAP {\bf 1010}, 008 (2010), arXiv:1005.2175.

\bibitem{Hunt:2013bha}
P.~Hunt and S.~Sarkar,
\newblock JCAP {\bf 1401}, 025 (2014), arXiv:1308.2317.

\bibitem{Joy:2007na}
M.~Joy, V.~Sahni, and A.~A. Starobinsky,
\newblock Phys.Rev. {\bf D77}, 023514 (2008), arXiv:0711.1585.

\bibitem{Joy:2008qd}
M.~Joy, A.~Shafieloo, V.~Sahni, and A.~A. Starobinsky,
\newblock JCAP {\bf 0906}, 028 (2009), arXiv:0807.3334.

\bibitem{Romano:2016gop}
A.~E. Romano, S.~Mooij, and M.~Sasaki,
\newblock Phys. Lett. {\bf B761}, 119 (2016), arXiv:1606.04906.

\bibitem{Mooij:2015cxa}
S.~Mooij, G.~A. Palma, G.~Panotopoulos, and A.~Soto,
\newblock JCAP {\bf 1510}, 062 (2015), arXiv:1507.08481,
\newblock [Erratum: JCAP1602,no.02,E01(2016)].

\bibitem{Creminelli:2012ed}
P.~Creminelli, J.~Noreña, and M.~Simonović,
\newblock JCAP {\bf 1207}, 052 (2012), arXiv:1203.4595.

\bibitem{m}
J.~M. Maldacena,
\newblock JHEP {\bf 0305}, 013 (2003), arXiv:astro-ph/0210603.

\bibitem{Cheung:2007st}
C.~Cheung, P.~Creminelli, A.~L. Fitzpatrick, J.~Kaplan, and L.~Senatore,
\newblock JHEP {\bf 03}, 014 (2008), arXiv:0709.0293.

\bibitem{Langlois:2010xc}
D.~Langlois,
\newblock Lect. Notes Phys. {\bf 800}, 1 (2010), arXiv:1001.5259.

\bibitem{Bunch:1978yq}
T.~Bunch and P.~Davies,
\newblock Proc.Roy.Soc.Lond. {\bf A360}, 117 (1978).

\bibitem{Noumi:2014zqa}
T.~Noumi and M.~Yamaguchi,
\newblock (2014), arXiv:1403.6065.

\bibitem{Planck:2013jfk}
Planck Collaboration, P.~Ade {\em et~al.},
\newblock (2013), arXiv:1303.5082.

\bibitem{xc}
X.~Chen,
\newblock Adv.Astron. {\bf 2010}, 638979 (2010), arXiv:1002.1416.

\bibitem{hann}
S.~Hannestad, T.~Haugbolle, P.~R. Jarnhus, and M.~S. Sloth,
\newblock JCAP {\bf 1006}, 001 (2010), arXiv:0912.3527.

\bibitem{Cai:2016ldn}
Y.~Cai, Y.-T. Wang, and Y.-S. Piao,
\newblock Phys. Rev. {\bf D94}, 043002 (2016), arXiv:1602.05431.

\bibitem{GallegoCadavid:2017bzb}
A.~Gallego~Cadavid, A.~E. Romano, and M.~Sasaki,
\newblock JCAP {\bf 1805}, 068 (2018), arXiv:1703.04621.

\bibitem{Wands:1998yp}
D.~Wands,
\newblock Phys. Rev. {\bf D60}, 023507 (1999), arXiv:gr-qc/9809062.

\bibitem{Boyle:2004gv}
L.~A. Boyle, P.~J. Steinhardt, and N.~Turok,
\newblock Phys. Rev. {\bf D70}, 023504 (2004), arXiv:hep-th/0403026.

\bibitem{Wang:2013eqj}
Y.~Wang,
\newblock Commun. Theor. Phys. {\bf 62}, 109 (2014), arXiv:1303.1523.

\bibitem{2009arXiv0902.1529K}
W.~H. {Kinney},
\newblock ArXiv e-prints , arXiv:0902.1529 (2009), arXiv:0902.1529.

\bibitem{Ashoorioon:2006wc}
A.~Ashoorioon and A.~Krause,
\newblock (2006), arXiv:hep-th/0607001.

\bibitem{Cai:2015xla}
Y.-F. Cai, E.~G.~M. Ferreira, B.~Hu, and J.~Quintin,
\newblock Phys. Rev. {\bf D92}, 121303 (2015), arXiv:1507.05619.

\bibitem{Creminelli:2004yq}
P.~Creminelli and M.~Zaldarriaga,
\newblock JCAP {\bf 0410}, 006 (2004), arXiv:astro-ph/0407059.

\bibitem{Adshead:2013zfa}
P.~Adshead, W.~Hu, and V.~Miranda,
\newblock Phys. Rev. {\bf D88}, 023507 (2013), arXiv:1303.7004.

\bibitem{Passaglia:2018afq}
S.~Passaglia and W.~Hu,
\newblock Phys. Rev. {\bf D98}, 023526 (2018), arXiv:1804.07741.

\bibitem{Abazajian:2016yjj}
CMB-S4, K.~N. Abazajian {\em et~al.},
\newblock (2016), arXiv:1610.02743.

\bibitem{2017arXiv170200786A}
P.~{Amaro-Seoane} {\em et~al.},
\newblock arXiv e-prints  (2017), arXiv:1702.00786.

\bibitem{Seoane:2013qna}
eLISA, P.~A. Seoane {\em et~al.},
\newblock (2013), arXiv:1305.5720.

\end{thebibliography}
\bibliographystyle{h-physrev4}
\end{document}